\documentclass[journal=nalefd,manuscript=letter]{achemso}

\usepackage[utf8]{inputenc}
\usepackage{graphicx}
\usepackage{amsmath,amsfonts,amssymb}

\title{Configurational control of photon emission from a molecular dimer}

\author{Maximilian Kögler}
\affiliation{Institut für Physik, Technische Universität Ilmenau, D-98693 Ilmenau, Germany}
\email{max.koegler@tu-ilmenau.de}

\author{Nicolas Néel}
\affiliation{Institut für Physik, Technische Universität Ilmenau, D-98693 Ilmenau, Germany}

\author{Jörg Kröger}
\affiliation{Institut für Physik, Technische Universität Ilmenau, D-98693 Ilmenau, Germany}
\email{joerg.kroeger@tu-ilmenau.de}

\begin{document}

\maketitle

\begin{abstract}
Tin-phthalocyanine molecules adsorbed on a NaCl ultrathin film on Au(111) exhibit electrofluorescence excited by a current across a scanning tunneling microscope junction.
Exploring the dependence of the molecular monomer photon yield on the injected current evidences the one-electron excitation process underlying the neutral-exciton luminescence.
Photon spectra of the monomer exhibit vibrational progression and hot luminescence, while the dimer electrofluorescence spectroscopic fine structure results from the coupling of the adjacent optical transition dipoles.
The photon yield of the dimer is significantly altered upon changing the configurational state of one of the two molecules.
In one of the bistable configurations light emission is amplified compared to the monomer, and it is reduced in the other.
\begin{tocentry}
\centering
\includegraphics[width=\linewidth]{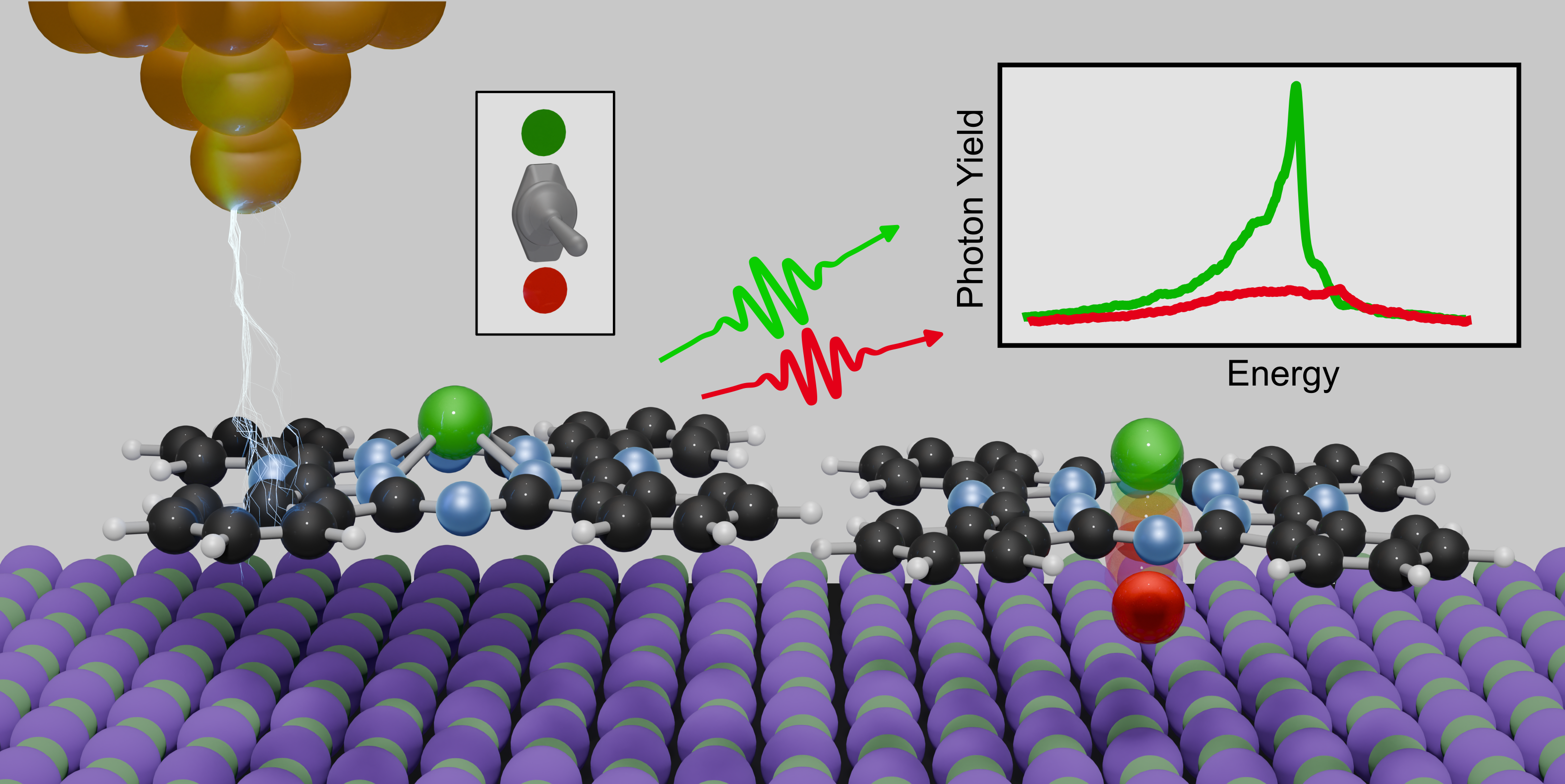}
\label{TOC}
\end{tocentry}
\end{abstract}


Research devoted to quantum photon emitters, for instance semiconductor quantum dots \cite{nature_406_968,science_295_102,nature_419_594}, atoms \cite{prl_39_691,prl_89_067901}, color centers \cite{ol_25_1294,natphoton_6_299,scirep_5_12882}, and molecules \cite{prl_69_1516,nature_407_491,prl_94_223602,prl_104_123605,natcommun_3_628,prl_108_093601}, is stimulated by the rapid progression of quantum-technological applications, e.\,g., cryptology and information processing \cite{nature_484_78,prx_7_031024,prl_117_243601,prl_121_123606,prl_122_093601,nature_453_1023,natphoton_10_631}, quantum computing \cite{prxquantum_6_020304}, the simulation of quantum mechanics \cite{natphys_8_285}, quantum sensing \cite{natphys_18_15} and metrology \cite{aqt_3_1900083}.
Likewise, fundamental aspects, for example the coupling of the exciton underlying the photon emission to the optical cavity for enhanced photon yield \cite{natcommun_8_15225,prl_119_013901,nl_18_3407,natmater_20_1615} or the radiative correlation of a group of quantum emitters giving rise to the coherent superposition of the individual emitter states \cite{prl_30_309,prl_36_1035,prl_38_764,prl_76_2049,pnas_94_10630,science_298_385,natphys_3_106,nature_531_623,prl_117_073002,natcommun_8_15225,nature_563_671,jpcc_123_21281,prl_122_233901}, belong to the highly topical developments in the field.

The scanning tunneling microscope (STM) plays a particularly appealing role in exploring quantum optical effects for several reasons.
To thoroughly study structural, electronic, and optical properties of quantum emitters, such as single molecules, the atomic scale, which is the resolution realm of an STM \cite{prl_49_57}, must be accessible.
In addition, the atom-by-atom manipulation of matter by using the STM tip \cite{nature_325_419,nature_344_524} is advantageous to fabricate artificial assemblies of atomic or molecular emitters.
Moreover, the spectroscopy of STM-induced light emission (STML) was demonstrated for molecular crystals \cite{science_262_1425} and single molecules \cite{science_299_542}, which opened the avenue to exploring quantum optics at the single-atom and single-molecule level \cite{ssr_65_129,chemrev_117_5174}.

Of particular interest to the work presented here is the amplification of photon emission reported for Zn-phthalocyanine (Zn-Pc) dimers \cite{nature_531_623} and for chains of up to twelve Zn-Pc molecules \cite{prl_122_233901} on NaCl-covered Ag(100) compared to the photon yield of the monomer.
In both cases, the coherent dipole-dipole coupling between the molecules in close proximity induced a collective state underlying the observed photon yield enhancement.
The report presented here was mainly motivated by identifying a molecular assembly that allows the external control of light amplification at the single-molecule level.
The use of Sn-Pc monomers and dimers on NaCl-covered Au(111) proved successful to this end.
The Sn-Pc molecule offers the reversible transfer of its central Sn atom through the macrocycle resulting in two bistable configurations with the Sn atom above (Sn-Pc-$u$) and beneath (Sn-Pc-$d$) the molecular backbone \cite{jacs_131_3639}, which are achieved by charge injection from the STM tip.
An Sn-Pc dimer with the $uu$ combination of configurations gives rise to a larger photon yield than observed from the Sn-Pc-$u$ monomer, while light emission of the $ud$ combination is nearly suppressed.
The key finding is therefore the reversible switch between a bright and a dark optical state of an Sn-Pc dimer.
In addition, since STML of Sn-Pc is reported here for the first time, the present work thoroughly characterizes the electroluminescence spectra of the monomer and the dimer.
The photon emission is due to a neutral molecular exciton and represents a one-electron excitation.
While the STML spectrum of an Sn-Pc-$u$ monomer exhibits fine structure due to vibrational progression and hot luminescence, the Sn-Pc-$u$ dimer gives rise to emission characteristics that are compatible with coupled molecular transition dipole moments. 

\begin{figure}
\centering
\includegraphics[width=\textwidth]{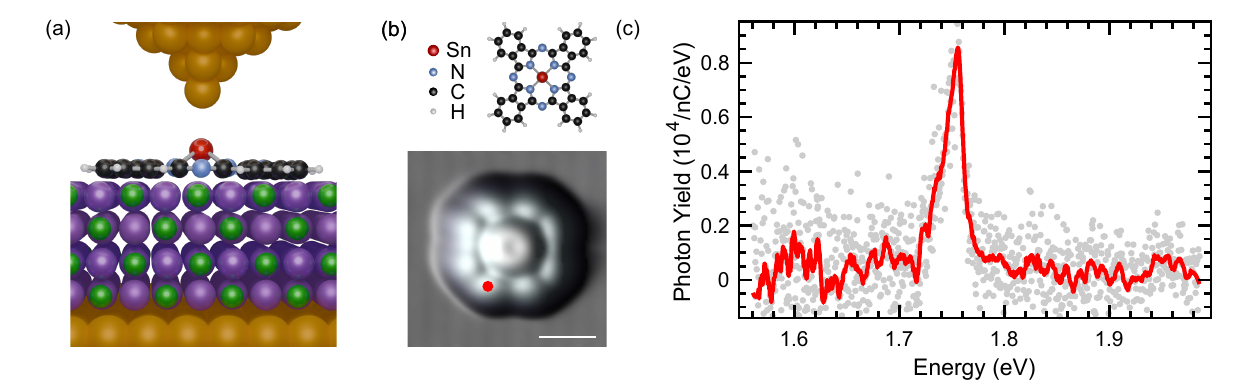}
\caption{Electroluminescence spectroscopy of an Sn-Pc-$u$ monomer.
(a) Sketch of the junction geometry including an Au-covered tip and the top surface layer (yellow) of Au(111) as well as a single Sn-Pc-$u$ molecule (H: white, C: black, N: blue, Sn: red) residing atop $4$ atomic NaCl (Na: green, Cl: violet) layers.
(b) Top: top view of the relaxed vacuum structure of Sn-Pc. 
Bottom: constant-current STM image of a single Sn-Pc-$u$ molecule on NaCl (sample voltage: $1.9\,\text{V}$, current: $20\,\text{pA}$, the gray scale covers apparent heights ranging from $0\,\text{pm}$ (dark) to $695\,\text{pm}$ (bright))\@.
The scale bar indicates $1\,\text{nm}$. 
(c) Wide range STML spectrum (dots) of Sn-Pc-$u$ acquired at $2.5\,\text{V}$, $80\,\text{pA}$ for $180\,\text{s}$ above the position indicated by the dot in (b)\@. 
The solid line represents smoothed data.}
\label{fig1}
\end{figure}

Figure \ref{fig1}a illustrates the basic STM junction geometry of the STML experiments comprising an Au-covered Ag tip and an Sn-Pc molecule residing atop an ultrathin NaCl film on Au(111) (details are presented in the Supporting Information, Figure S1)\@.
Three and four atomic layers of NaCl with apparent heights (measured at $1.9\,\text{V}$) of about $430\,\text{pm}$ and $580\,\text{pm}$, respectively, resulted from the preparation protocol (Methods), which is in agreement with previous work \cite{jpcm_24_475507}.
Unless otherwise stated, the presented data were obtained on $4$-layer thin NaCl islands.

A typical STM image of a single Sn-Pc-$u$ is presented in Figure \ref{fig1}b.
The strongest contribution to the image is due to the Sn atom, which in this molecular configuration points toward the tip.
The Sn-associated protrusion exhibits a shallow central depression, which is assigned to the orbital texture of the molecule, as reported earlier \cite{acie_48_1261,jacs_131_3639}.
The outer regions of the four cloverlike isoindole moieties appear as two split lobes each, which was observed for Sn-Pc on molecular spacers, too, and associated with the charge density pattern of the lowest unoccupied molecular orbital (LUMO) \cite{jacs_131_3639}.

Figure \ref{fig1}c depicts the photon yield spectrum of Sn-Pc-$u$ in a wide range of photon energies.
Details of optical data processing are described in the Supporting Information (Figure S2)\@.
The monomer spectrum was obtained by parking the tip above one of the split lobes (dot in Figure \ref{fig1}b) at positive voltage.
Elevated negative voltages turned out to move Sn-Pc-$u$ and, thus, were not suitable for STML\@.
The emission peak at $1.75\,\text{eV}$ is consistent with the previously determined absorbance maximum of Sn-Pc in a benzene solution of $1.78\,\text{eV}$ \cite{prb_93_115418, jpcm_31_134004}.
The discrepancy in emission and absorbance maxima is likely due to the different dielectric environments in the two cases.
In accordance with theoretical studies \cite{daltontrans_41_7141}, the Sn-Pc absorption was assigned to the Q-band, which mainly involves the electronic excitation from the highest occupied molecular orbital (HOMO) to the LUMO\@.
Therefore, the Sn-Pc-$u$ photon emission line visible in Figure \ref{fig1}c can reasonably be associated with the neutral Q-exciton of the molecule.
Electrofluorescence spectra of the Sn-Pc-$d$ monomer were hampered due to its unintentional conversion to Sn-Pc-$u$ at the tunneling gap parameters required for a decent photon signal.

\begin{figure}
\centering
\includegraphics[width=\textwidth]{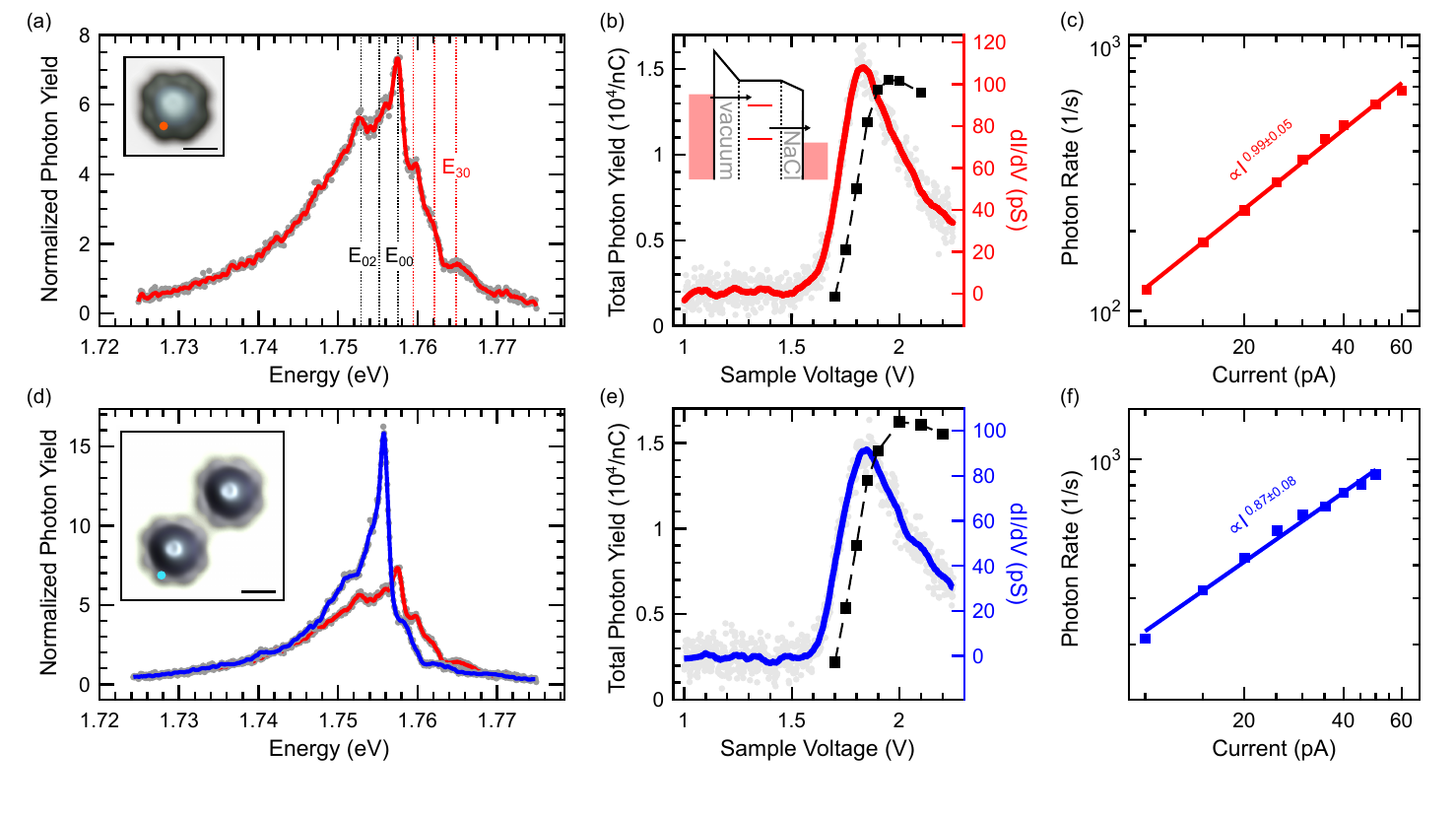}
\caption{Photon emission characteristics of Sn-Pc-$u$ monomers and dimers.
(a) Spectral photon yield of Sn-Pc-$u$ recorded at $1.9\,\text{V}$, $50\,\text{pA}$ for $180\,\text{s}$. 
The data are normalized by the spectrum of the tip-induced plasmon acquired above the bare NaCl island with the same tip.
Inset: constant-height STM image of Sn-Pc-$u$ ($1.9\,\text{V}$, the gray scale encodes currents ranging from $10\,\text{pA}$ (dark) to $100\,\text{pA}$ (bright), scale bar: $1\,\text{nm}$)\@.
The dot indicates the tip position for luminescence spectroscopy.
(b) Constant-height spectrum of $\text{d}I/\text{d}V$ (dots) recorded above the same position as the light spectrum (feedback loop parameters: $1.9\,\text{V}$, $20\,\text{pA}$)\@.
The solid line represents smoothed data.
Squares depict the integrated spectral photon yield of (a) with the dashed line as a guide to the eye. 
Inset: illustration of alignment of tip and sample electron chemical potentials with molecular frontier orbitals for the resonant excitation of the molecule by tunneling electrons.
Arrows indicate tunneling of electrons from the tip across the vacuum barrier to the LUMO and from the HOMO across the NaCl barrier to the sample.
(c) Evolution of photon rate (squares) with the junction current at $1.9\,\text{V}$\@.
The solid line is a power law fit to the data.
(d)--(f) As (a)--(c) for an Sn-Pc-$u$ dimer.
In (d), the acquisition time for the dimer spectrum is $240\,\text{s}$ and a monomer spectrum (red) recorded with the same tip as the dimer spectrum is added for comparison. 
Inset to (d): constant-height STM image of an artificially assembled Sn-Pc-$u$ dimer ($1.9\,\text{V}$, the gray scale encodes currents ranging from $10\,\text{pA}$ (dark) to $75\,\text{pA}$ (bright), scale bar: $1\,\text{nm}$)\@.}
\label{fig2}
\end{figure}

In Figure \ref{fig2}, electrofluorescence spectroscopy with enhanced optical resolution of Sn-Pc-$u$ monomers and dimers is analyzed in detail.
The emission spectrum of a single Sn-Pc-$u$ molecule is shown in Figure \ref{fig2}a.
Besides the principal transition from the electronic excited ($S_1$) to the electronic ground ($S_0$) state with energy $E_{00}=1.758\,\text{eV}$), spectroscopic fine structure is observed both at energies below ($E_{i0}$) and above ($E_{0i}$) the $S_1\rightarrow S_0$ transition (dashed lines)\@. 
While the emission features for $E_{i0}<E_{00}$ are attributed to vibrational progression, the group of emission lines with $E_{0i}>E_{00}$ is assigned to hot luminescence.
The indices in $E_{\nu\nu'}$ denote the vibrational states of $S_0$ ($\nu$) and $S_1$ ($\nu'$) with $\nu=0$ ($\nu'=0$) reflecting the vibrational ground state of $S_0$ ($S_1$)\@.
A fit to the data using a superposition of equidistant Lorentzian line shapes (Supporting Information, Figure S3) performed separately for $E<E_{00}$ and  $E>E_{00}$ reproduces the emission spectrum and yields vibrational energy quanta of $2.5\,\text{meV}$ and $3.0\,\text{meV}$ for $S_0$ and $S_1$, respectively. 
Such low-energy vibrational excitations can be assigned to librons, which represent frustrated rotations of the molecule and were observed for other phthalocyanine molecules, too \cite{natcommun_13_6008}. 

Figure \ref{fig2}b compares the LUMO-related $\text{d}I/\text{d}V$ signature (dots) with the total photon yield (squares) of the Sn-Pc-$u$ monomer, which represents the integrated spectral photon yield (total photon yield) in the energy interval $1.7\,\text{eV}\leq E\leq 1.8\,\text{eV}$\@.
The LUMO-associated signal increases with increasing voltage and reaches its maximum at $1.8\,\text{V}$\@.
The total photon yield follows the same trend, albeit shifted by $\approx 0.2\,\text{V}$ to higher voltages.
Consequently, for Sn-Pc-$u$ photon emission the alignment of the tip electron chemical potential with the LUMO is required (inset to Figure \ref{fig2}b)\@.
Moreover, assuming the HOMO alignment with the electron chemical potential of the sample, the Q-exciton of Sn-Pc-$u$ is formed by an electron attached to the LUMO and a hole transferred to the HOMO, as observed for electroluminescence of a variety of molecules \cite{science_299_542,prb_77_205430,prl_105_217402,nature_531_623,prl_112_047403,natcommun_8_580}.
This assumption is in line with $\text{d}I/\text{d}V$ spectroscopy of the molecular frontier orbitals (Supporting Information, Figure S4) after considering the voltage drop across the NaCl film and the deviation of the sought-after LUMO (HOMO) from the measured negative (positive) ion resonance \cite{jpcm_20_184001}.
The attachment of one electron and one hole to the frontier orbital levels also reflects the photon emission of the molecule in its neutral state.
By exploring the current dependence of the photon rate, which is the product of the total photon yield and the current a power law behavior $I^\alpha$ with $\alpha=0.99\pm 0.05$ is found (Figure \ref{fig2}c)\@.
The near-unity exponent evidences a one-electron process in the electron-to-photon conversion.

Dimers were fabricated by pushing one monomer with the tip to a second Sn-Pc-$u$ molecule, which was achieved with feedback loop parameters of $-3.1\,\text{V}$ and $20\,\text{pA}$ and the tip placed atop an isoindole group of the pushed monomer (Supporting Information, Figure S5)\@.
Figure \ref{fig2}d shows the electrofluorescence spectrum of an Sn-Pc dimer in the $uu$ configuration, where the dimer partners are separated by $2.3\,\text{nm}$ (measured between the Sn centers, inset to Figure \ref{fig2}d)\@.
The spectral centroid is redshifted by $\approx 2\,\text{meV}$ with respect to the monomer spectrum and the extent of the redshift increases with decreasing intermolecular separation (Supporting Information, Figure S6)\@. 
In addition, the maximum photon yield of the dimer exceeds the one of the monomer by nearly a factor of $2$, and at the same time the overall width of the spectrum is smaller than observed for the monomer.
Furthermore, the spectroscopic fine structure visible in the $uu$ dimer spectrum is of different nature and will be discussed below.
The variation of the $\text{d}I/\text{d}V$ signature of the LUMO is imprinted on the voltage evolution of the total photon yield (Figure \ref{fig2}e), in a similar manner as observed for the monomer.
Consequently, alignment of the electrode chemical potentials with the frontier orbitals is required for the dimer photon emission, which also occurs in the neutral state of the $uu$ hybrid.
A power law behavior of the photon rate with the tunneling current (Figure \ref{fig2}f) is likewise observed with an exponent of $\alpha=0.87\pm 0.08$ that within the uncertainty margins is similar to the exponent observed for the monomer.

\begin{figure}
\centering
\includegraphics[width=\textwidth]{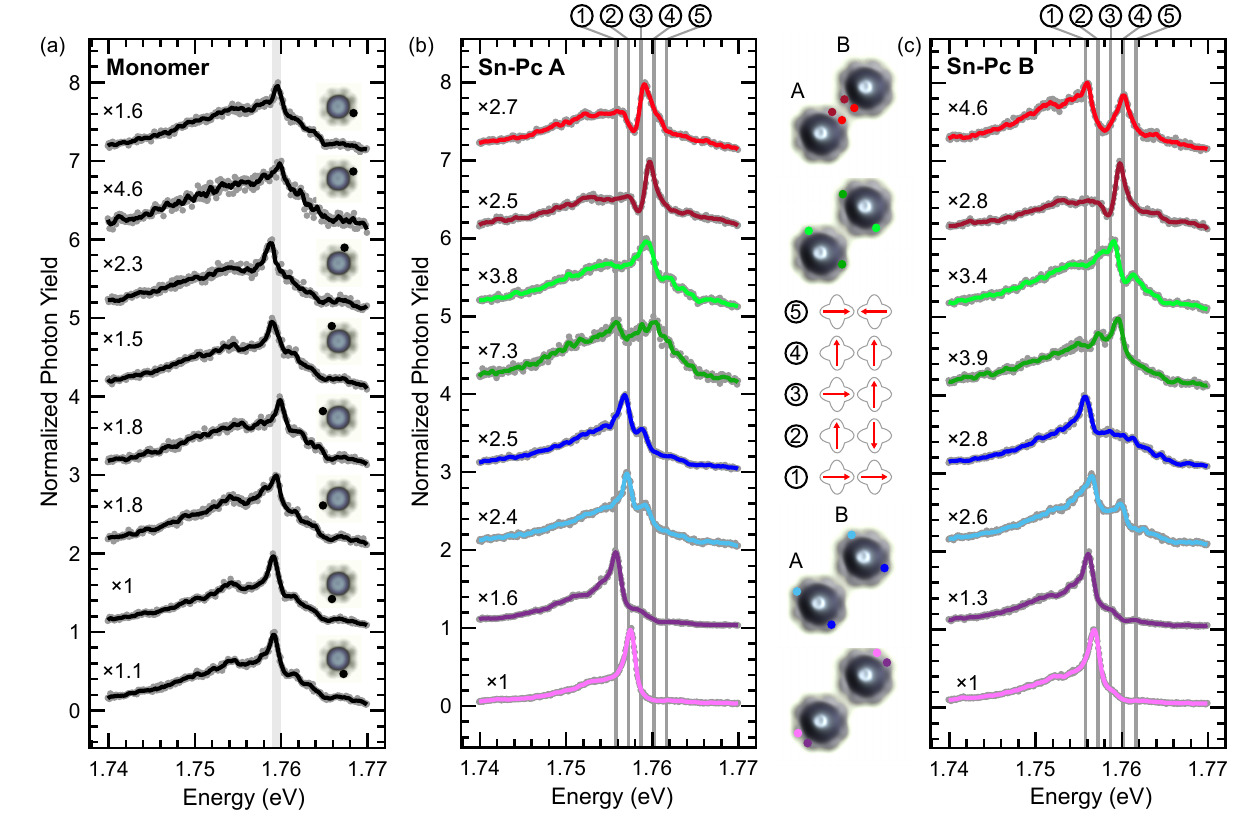}
\caption{Spatially resolved electroluminescence spectra of the Sn-Pc-$u$ monomer and $uu$ dimer.
(a) Site dependence of the normalized monomer spectral photon yield ($1.9\,\text{V}$, $50\,\text{pA}$, $120\,\text{s}$)\@. 
The position of spectroscopy is marked by the dot in the constant-height STM images ($1.9\,\text{V}$, $2.3\,\text{nm}\times 2.3\,\text{nm}$, the gray scale encodes currents ranging from $2\,\text{pA}$ (dark) to $30\,\text{pA}$ (bright))\@.
The vertical stripe has a width of $1\,\text{meV}$ and marks the $S_1\rightarrow S_0$ transition in all spectra.
(b),(c) Site dependence of the normalized $uu$ dimer spectral photon yield ($1.9\,\text{V}$, $50\,\text{pA}$, $120 \,\text{s}$) atop the indicated (dots) positions.
The position of spectroscopy is marked by the dot in the constant-height STM images ($1.9\,\text{V}$, $5\,\text{nm}\times 5\,\text{nm}$, the gray scale encodes currents ranging from $10\,\text{pA}$ (dark) to $75\,\text{pA}$ (bright)), where the individual monomer molecules of the dimer are labeled $A$ and $B$\@.
The vertical lines $1$--$5$ indicate the dipole-coupled exciton states (see illustrations between (b) and (c)) underlying the fine structure.
The spectra in (a)--(c) are normalized to their corresponding maximum and vertically offset.}
\label{fig3}
\end{figure}

While the voltage dependence of the total photon yield and the current dependence of the photon rate are similar for the monomer and dimer electroluminescence, the spatial dependence of monomer and dimer light emission is markedly different (Figure \ref{fig3})\@.
The explanation of these differences includes the revelation of the different nature of the dimer luminescence fine structure.
To explore the spatial emission characteristics of the Sn-Pc-$u$ monomer (Figure \ref{fig3}a) the tip was parked above each of the $8$ lobes.
At each lobe, the line shapes of the monomer spectra are similar and are dominated by the emission peak due to the $S_1\rightarrow S_0$ transition (gray stripe in Figure \ref{fig3}a)\@.
Its position changes by at most $1\,\text{meV}$\@.
These changes may be due to the photonic Lamb shift \cite{natcommun_8_15225} caused by the coupling of the nanocavity plasmon and the molecular transition dipoles.
The spatial magnitude of the photon yield shows a marked variation depending on the lobe, which does not reflect the $C_{2v}$ symmetry of the molecule.
While a clear-cut rationale for this observation is missing to date, the exact lobe position atop the atomic NaCl lattice may be important.
Indeed, extrapolating the atomically resolved NaCl lattice beneath the Sn-Pc-$u$ monomer reveals that the isoindole moieties do not share equivalent adsorption sites (Supporting Information, Figure S7)\@.

In striking contrast to the observations for the monomer, the dimer spectral photon yield exhibits pronounced peaks in an energy interval of $6\,\text{meV}$ (vertical lines) whose signal strength varies markedly depending on the position of spectroscopy (Figures \ref{fig3}b,c)\@.
In total, $5$ major transition peaks can be identified.
At the same time, the spatially resolved $\text{d}I/\text{d}V$ spectra of the dimer (Figure \ref{fig2}e and Supporting Information, Figure S4) are similar to the monomer data, which reflects the spatially unaltered electronic level alignment for photon emission.
Therefore, the $uu$ dimer luminescence fine structure is rationalized in terms of the coherent intermolecular transition dipole coupling (TDC), which was previously reported for Zn-Pc oligomers \cite{nature_531_623}. 
Due to TDC, the molecular dimer exhibits two single-exciton states for each of the two transition dipole moments with wave functions $\Psi^\pm=(\psi_{1\text{e}}\psi_{2\text{g}}\pm\psi_{1\text{g}}\psi_{2\text{e}})/\sqrt{2}$ where $\psi_{k\text{g}}$, $\psi_{k\text{e}}$ denote the ground (g) and excited (e) exciton states of the monomer $k=1,2$\@.
The associated dimer exciton energies are $E^\pm=E_{00}+\Delta\pm\vert J\vert$ with $E_{00}$ the monomer exciton energy, $\Delta$ the van der Waals energy difference between the dimer ground and excited states, and $\vert J\vert$ the coupling strength between the monomer transition dipoles $\boldsymbol{\mu}_k$.
In the point-dipole approximation \cite{pac_11_371}, $J=J_0\left[\boldsymbol{\mu}_1\cdot\boldsymbol{\mu}_2-3(\boldsymbol{\mu}_1\cdot\hat{\boldsymbol{r}})(\boldsymbol{\mu}_2\cdot\hat{\boldsymbol{r}})\right]$ with $J_0=1/(4\pi\varepsilon_0r^3)$, $\varepsilon_0$ the vacuum permittivity, $r$ the center-to-center distance of the monomers, and $\hat{\boldsymbol{r}}=\boldsymbol{r}/r$ the associated unit distance vector.
For calculating $J$, different orientations of $\boldsymbol{\mu}_k$ are considered.
While for planar Zn-Pc the transition dipole moments are oriented parallel to the molecular plane \cite{nature_531_623}, the three-dimensional shuttlecock geometry of Sn-Pc (Figure \ref{fig1}a) \cite{daltontrans_41_7141,aipadv_2_041402,jacs_131_3639} possibly gives rise to an out-of-plane orientation of $\boldsymbol{\mu}_k$.
However, the monomer exciton involves the HOMO and LUMO, which are mainly confined to the almost planar isoindole moieties of Sn-Pc \cite{daltontrans_41_7141,prb_93_115418,jpcm_31_134004}.
Therefore, planar transition dipoles are assumed for Sn-Pc as well.
Defining then directions of $\boldsymbol{\mu}_k$ as $\uparrow$, $\downarrow$ perpendicular to $\hat{\boldsymbol{r}}$ and as $\leftarrow$, $\rightarrow$ parallel to $\hat{\boldsymbol{r}}$, $5$ TDC modes (Figures \ref{fig3}b,c) $\rightarrow\rightarrow$ (1), $\uparrow\downarrow$ (2), $\rightarrow\uparrow$ (3), $\uparrow\uparrow$ (4), $\rightarrow\leftarrow$ (5) with energies $-2\mu^2J_0$, $-\mu^2J_0$, $0$, $\mu^2J_0$, $2\mu^2J_0$, respectively, are expected for $\mu=\vert\boldsymbol{\mu}_1\vert=\vert\boldsymbol{\mu}_2\vert$\@.
With these assumptions, $\mu^2J_0$ can be determined from the experimental data as the average distance between adjacent spectral photon yield peaks (vertical lines in Figures \ref{fig3}b,c), which results in $(1.5\pm 0.2)\,\text{meV}$ and which is comparable with the smallest splitting reported for Zn-Pc ($(3.0\pm 1.5)\,\text{meV}$) \cite{nature_531_623}.
In a few spectra, a broad spectral photon yield peak was found between two energies defined by neighboring vertical lines, which is likely caused by the superposition of two adjacent peaks.

\begin{figure}
\centering
\includegraphics[width=\textwidth]{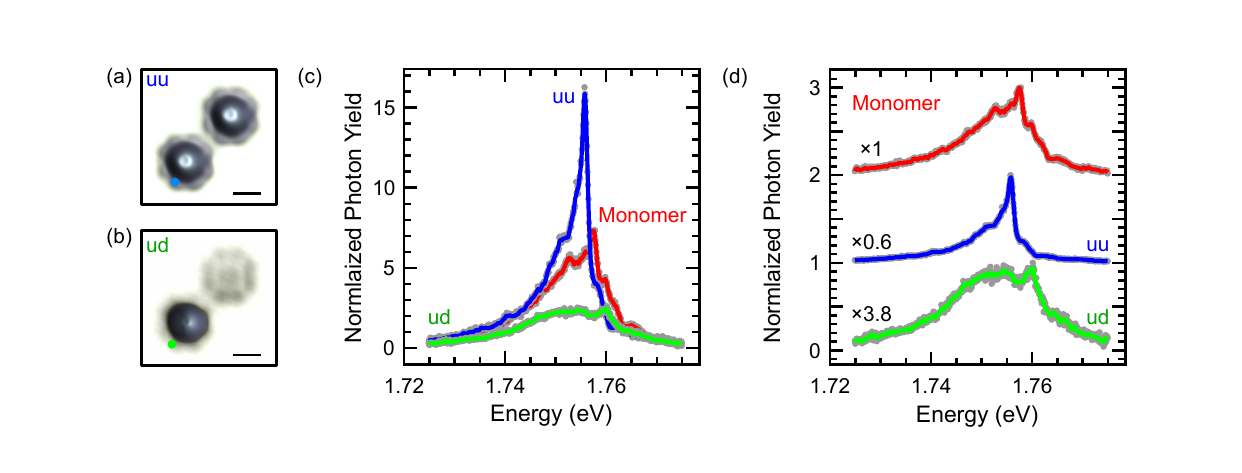}
\caption{Electrofluorescence of $uu$ and $ud$ Sn-Pc dimers.
(a),(b) Constant-height STM image of a $uu$ (a) and a $ud$ (b) dimer ($1.9\,\text{V}$, the gray scale encodes currents from (a) $10\,\text{pA}$, (b) $2\,\text{pA}$ (dark) to (a) $75\,\text{pA}$, (b) $25\,\text{pA}$ (bright), scale bar: $1\,\text{nm}$)\@. 
(c) Normalized spectral photon yield of an Sn-Pc-$u$ monomer and Sn-Pc dimers in the $uu$ and $ud$ configuration ($1.9\,\text{V}$, $50\,\text{pA}$, $240\,\text{s}$)\@. 
All spectra were recorded with the same tip above the positions indicated by the dots in (a),(b)\@.
They are normalized by the tip plasmon luminescence spectrum atop bare NaCl.
(d) Same as (c) with the spectra normalized by their respective maximum and vertically offset.}
\label{fig4}
\end{figure}

Before addressing the amplification of the photon yield in the TDC model, the electroluminescence of the dimer in the $ud$ configuration is discussed.
Stimulated by electron injection, the Sn atom can reversibly be transferred across the molecular macrocycle, as originally demonstrated for molecular decoupling layers \cite{jacs_131_3639} and later for various substrates \cite{jpcm_24_394004,jpcc_119_15716,acsnano_19_7231}. 
Placing the tip above the central Sn atom of an Sn-Pc-$u$ molecule at negative voltage leads to the $u\rightarrow d$ conversion, while the $d\rightarrow u$ conversion is achieved for positive voltage (Supporting Information, Figure S5)\@.
Unlike for Sn-Pc-$u$, the isoindole groups protrude towards the tip for Sn-Pc-$d$, and the center is imaged as a depression. 
Figure \ref{fig4} compares electroluminescence spectroscopy for an Sn-Pc dimer in the $uu$ (Figure \ref{fig4}a) and in the $ud$ (Figure \ref{fig4}b) configuration.
In order to avoid the unintentional $d\rightarrow u$ conversion at the elevated positive voltages required for electroluminescence spectroscopy, all spectra (Figure \ref{fig4}c) were recorded above the Sn-Pc-$u$ molecule.
While the $uu$ configuration of the dimer exceeds the spectral photon yield of the monomer by a factor of nearly $2$, the $ud$ configuration gives rise to a maximum spectral photon yield that falls below the monomer emission by a factor of almost $4$, that is, the maximum spectral photon yields of the $uu$ and $ud$ dimer are more than a factor $6$ apart.
Additionally, spectroscopic fine structure is hardly observed in the emission spectrum of the $ud$ dimer.
Rather, the shape of the $ud$ dimer spectrum resembles the broad background of the monomer spectrum with an indentation at the position of the monomer principal $S_1\rightarrow S_0$ emission.
The sharper feature at $1.76\,\text{eV}$ may be due to the dipole-coupled $\uparrow\uparrow$ exciton mode of the TDC model (Figure \ref{fig4}d)\@. 

The remainder of this article is devoted to the tentative explanation of the markedly different photon yield of the $uu$ and $ud$ Sn-Pc dimers.
The strong amplification of electroluminescence of the $uu$ dimer relative to the $u$ monomer is reminiscent of superradiance, which describes a cooperative optical effect that results from the radiative correlation of quantum emitters \cite{pr_93_99} provided by intermolecular dipole-dipole coupling resulting in a quantum coherent collective state \cite{pr_93_99,science_325_1510,prl_116_163604}.
Superradiance was reported for Zn-Pc oligomers with STML \cite{prl_122_233901}.
In agreement with the findings reported here, with increasing length of the oligomer chain the photon yield was augmented, the width of the electroluminescence spectrum narrowed, and its centroid was redshifted.
This effect was particularly obvious for the transition from the Zn-Pc monomer to the dimer \cite{prl_122_233901}.
The TDC model used above for rationalizing the electroluminescence fine structure of the $uu$ dimer corroborates the superradiance scenario in that the dipole-coupled $\uparrow\uparrow$ exciton state is strongly enhanced for the $uu$ dimer compared to the monomer, which hints at the in-phase superposition of the $\uparrow$ transition dipole moments of the two individual monomers.
It is then tempting to associate the low photon yield of the $ud$ dimer with a subradiant state. 
However, the transition dipole moments of Sn-Pc-$d$ are expected to be the same as for Sn-Pc-$u$, and their coupling in the $ud$ dimer are unlikely to be altered compared to the $uu$ dimer \cite{daltontrans_41_7141, prb_93_115418}.
Indeed, the molecular orbitals underlying the exciton are the HOMO and the LUMO at the periphery of the macrocycle, which is nearly unaffected by the configurational switch.
The observed reduction in the photon yield of the $ud$ dimer could possibly be caused by different emission characteristics of the Sn-Pc-$d$ monomer in the following sense.
First, while for Sn-Pc-$d$ electroluminescence spectroscopy was experimentally not feasible (vide supra), indirect evidence for the suggested configurational changes in the emission characteristics are provided by a recent STML study of VO-Pc on NaCl-covered Au(111) \cite{arxiv_2508_11501}.
For the two molecular configurations defined by the O atom above and beneath the backbone of the adsorbed molecule, a difference in the principal emission peaks of $\approx 25\,\text{meV}$ was shown.
Second, a comparable difference was observed for the Q-exciton-associated electroluminescence of Pt-Pc ($1.945\,\text{eV}$) and Zn-Pc ($1.901\,\text{eV}$) on NaCl-covered Ag(111) \cite{natnanotechnol_17_729}.
Upon forming a heterodimer out of these molecules, the dimer principal emission peak was strongly redshifted to $1.887\,\text{eV}$\@.
A similar scenario could be applicable to the $ud$ dimer in the present case, too.

In conclusion, luminescence of a molecular dimer of Sn-Pc on NaCl-covered Au(111) can be mastered by changing the configuration of one of the constituent monomers.
The reversibility of the configurational switch allows to turn on and off the light emission at the single-molecule level.
The bright optical state is possibly due to superradiance as inferred from the comparison of the monomer STML spectrum, which exhibits vibrational progression and hot luminescence, with the dimer spectral data whose spectroscopic fine structure is due to excitonic states resulting from the coherent coupling of the monomer transition dipole moments.
The dark optical state is tentatively assigned to the redshifted principal emission caused by the presumably different emission characteristics of the bistable molecular configurations of the monomers.
The presented findings exemplify the external control of quantum coherent processes, that is, the on-demand presence of bright and dark optical states, which may be particularly relevant to the nanotechnology of quantum light sources.

\section{Methods}

The experiments were performed with an STM operated in ultrahigh vacuum ($10^{-9}\,\text{Pa}$) and at low temperature ($5\,\text{K}$)\@.
Surfaces of Au(111) were cleaned and prepared by Ar$^+$ ion bombardment and annealing.
Clean NaCl (purity: $\geq 99\,\%$) was sublimated at a temperature of $890\,\text{K}$ on clean Au(111) at room temperature. 
The NaCl flux was controlled by a quartz microbalance. 
After NaCl deposition for $100\,\text{s}$, the covered sample was annealed at $470\,\text{K}$ to yield a $50\,\%$ coverage of $3$ to $4$ atomic layers with of NaCl.
The solid phase of Sn-Pc molecules was sublimated from a ceramics crucible heated at $780\,\text{K}$. 
A coverage of approximately $20$ Sn-Pc molecules per $80\,\text{nm}\times 80\,\text{nm}$ was achieved after deposition for $2\,\text{min}$ onto the cold ($6\,\text{K}$) NaCl-covered sample surface. 
A chemically edged and subsequently Ar$^+$-bombarded Ag wire served as the tip material.
Field emission on and repeated indentations into the clean Au surface presumably presumably led to the coating of the tip apex with substrate material.
Single-atom transfer from the tip to the sample gave rise to particularly sharp and stable probes \cite{prl_94_126102,jpcm_20_223001,prl_102_086805,pccp_12_1022}.
Images were acquired in the constant-current and constant-height mode with the bias voltage applied to the sample and were further processed with the with the Nanotec Electronics WSxM software \cite{rsi_78_013705}.
Spectroscopy of $\text{d}I/\text{d}V$ proceeded via the sinusoidal modulation ($5\,\text{mV}$, $726\,\text{Hz}$) of the dc sample voltage and measuring the first harmonic of the ac current response of the tunneling junction with a lock-in amplifier.
For electroluminescence spectroscopy, a pair of two lenses, one placed inside and the other outside of the vacuum recipient, collects the light emitted from the tunneling junction (Supporting Information, Figure S1)\@.
High-resolution (wide-range) optical spectra were recorded with an optical grating of $1200$ ($300$) grooves per mm.

\section{Associated Content}

\subsection{Supporting Information} 
Supporting Information is available free of charge on the ACS Publication website at DOI: [hyperlink DOI] 

Electroluminescence detection and spectroscopy with a scanning tunneling microscope, electroluminescence data processing, analysis of the Sn-Pc-$u$ monomer electrofluorescence, spectroscopy of Sn-Pc monomer and dimer orbitals, fabrication of dimers and control of configurational molecular states, dependence of the dimer emission on the mutual monomer separation, adsorption of Sn-Pc on ultrathin NaCl films (PDF)

\subsection{Data availability statement} 
The data that support the findings of this article are not publicly available upon publication because it is not technically feasible and/or the cost of preparing, depositing, and hosting the data would be prohibitive within the terms of this research project. 
The data are available from the authors upon reasonable request. 

\subsection{Author contributions}
M.K. carried out the experiments and analyzed the data. 
M.K and J.K. wrote the manuscript. 
J.K. conceived the experiments.
All authors discussed the data, their interpretation, and commented on the final version of the manuscript.

\begin{acknowledgement}
Funding by the German Federal Ministry of Education and Research within the ''Forschungslabore Mikroelektronik Deutschland (ForLab)'' initiative is acknowledged. 
Discussions with Daniel Wegner on technical issues of light detection and with Lorenz Meyer, Marvin Kuß, Robert Henninger are appreciated.
\end{acknowledgement}

%\bibliography{ref.bib}
\providecommand{\latin}[1]{#1}
\makeatletter
\providecommand{\doi}
  {\begingroup\let\do\@makeother\dospecials
  \catcode`\{=1 \catcode`\}=2 \doi@aux}
\providecommand{\doi@aux}[1]{\endgroup\texttt{#1}}
\makeatother
\providecommand*\mcitethebibliography{\thebibliography}
\csname @ifundefined\endcsname{endmcitethebibliography}
  {\let\endmcitethebibliography\endthebibliography}{}

\end{document}

\maketitle

\section{Electroluminescence detection and spectroscopy with a scanning tunneling microscope} 

\begin{figure}
\centering
\includegraphics[width=\textwidth]{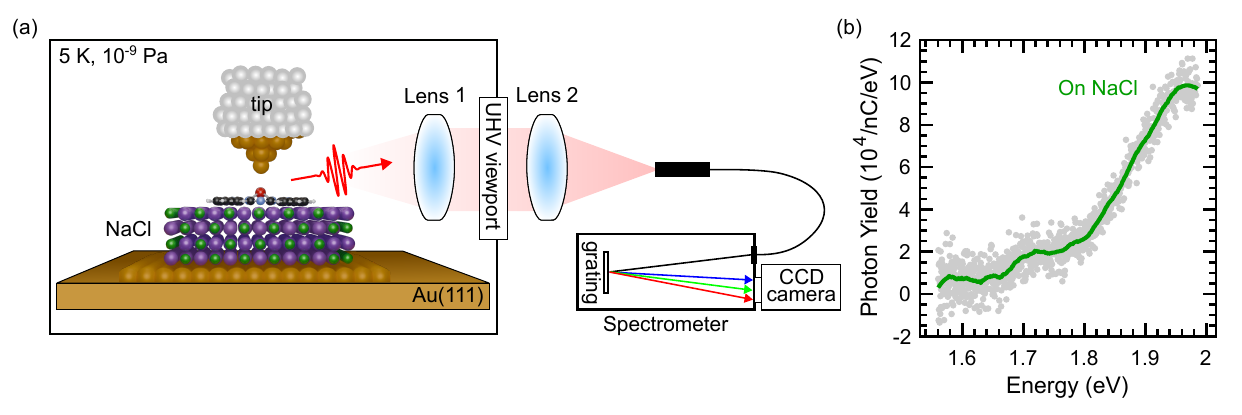}
\caption{Experimental setup for STML and spectroscopy of the tip-induced plasmon radiation.
(a) Sketch illustrating the setup for luminescence spectroscopy with an STM\@.
(b) Spectral photon yield of the NCP radiative decay above $3$ atomic layers of NaCl on Au(111) ($2.5\,\text{V}$, $20\,\text{pA}$, $180\,\text{s}$)\@.}
\label{figS1}
\end{figure}

The experimental setup used in this work is illustrated in Figure \ref{figS1}. 
The focal point of lens $1$ inside the vacuum recipient is the tunneling barrier between tip and sample.
The collected and collimated light traverses a viewport of the chamber and is focused by lens $2$ to an optical fiber. 
The fiber transports the light to the spectrometer.
For wide range (high resolution) spectra, an optical reflection grating with $300$ ($1200$) grooves per mm is used. 
The spatially dispersed light is detected by a charge-coupled device (CCD) camera, which is thermoelectrically cooled to $175\,\text{K}$. 

Figure \ref{figS1}b shows a typical spectral photon yield distribution of the nanocavity plasmon (NCP) radiative decay recorded above a $3$-layer NaCl island on Au(111)\@. 
A strong increase of the photon yield is observed for energies exceeding $1.8\,\text{eV}$\@. 
The small plateau between $1.7\,\text{eV}$ and $1.8\,\text{eV}$ marks the energy window of interest in the present studies. 
Below $1.7\,\text{eV}$, the light due to the NCP is essentially quenched. 

\section{Electroluminescence data processing} 

\begin{figure}
\centering
\includegraphics[width=\textwidth]{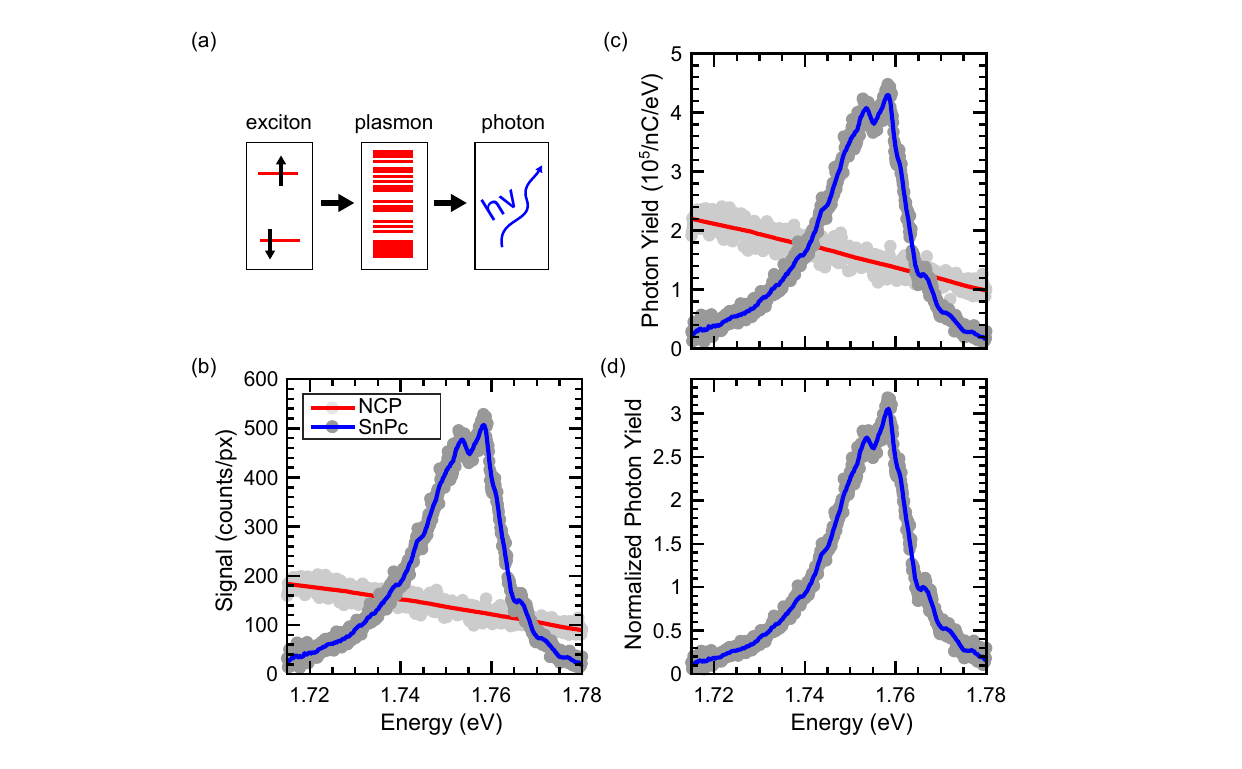}
\caption{Principles underlying the STML data processing.
(a) Sketch illustrating the radiative decay of a molecular exciton via the continuum of NCP modes. 
(b) Raw STML data acquired above an isoindole moiety of an Sn-Pc-$u$ molecule (dark gray dots) above $4$ atomic layers of NaCl on Au(111) ($1.9\,\text{V}$, $50\,\text{pA}$, $240\,\text{s}$) together with STML data of the NCP (light gray dots) above the pristine NaCl island ($1.9\,\text{V}$, $50\,\text{pA}$, $180\,\text{s}$), both corrected for the CCD background.
Solid lines are smoothed data. 
(c) Same as (b) using physical units. 
(d) Result of dividing the molecular by the NCP luminescence spectrum.}
\label{figS2}
\end{figure}

To understand the data processing it is useful to look at the photon emission mechanism in STML (Figure \ref{figS2}a)\@.
The quantum emitter is excited by charge injection, which gives rise to a molecular exciton that radiatively decays into a photon.
The tip-induced plasmon or the NCP represents an important ingredient for the photon emission.
It represents a collective charge oscillation in tip and sample, that is, a localized surface plasmon, which interacts with the tunneling electrons \cite{prl_41_1746,ssr_65_129,prl_68_3224}.
The light emission spectrum depicted in Figure \ref{figS1}b can be viewed as a map of the available plasmonic modes of the nanocavity.
The NCP density of states (DOS) at a specific energy determines the transition efficiency of an exciton with the same energy \cite{science_361_251, nl_18_3407}. 
Consequently, the NCP DOS can be viewed as an energy-dependent amplification factor for the exciton decay.
It is noteworthy that the tip-induced breaking of the in-plane momentum conservation is a necessary condition for the coupling of NCP and the photon. 

The red solid line in Figure \ref{figS2}b shows the raw NCP-associated luminescence obtained above $4$ atomic layers of NaCl on Au(111)\@. 
In the measured energy range, the NCP-related light signal is featureless, that is, it does not add any fine structure to the measured STML spectra above Sn-Pc.
To retrieve the genuine emission characteristics of the probed Sn-Pc-$u$ molecule, the electrofluorescence spectrum acquired above the molecule must be normalized by the NCP luminescence. 
First, the background of the CCD camera must be removed (Figure \ref{figS2}b)\@. 
Second, the raw STML signal, which gives the number of counts per CCD pixel (px), is translated into the physical unit (Figure \ref{figS2}c)\@.
To this end, the number of counts per CCD bin is divided by the associated energy interval and by the charge $I\tau$ ($\tau$: exposure time of the CCD array) transported across the tunneling junction.
Third, the electroluminescence data measured above the molecule is divided by the Savitzky-Golay-smoothed NCP spectral photon yield (Figure \ref{figS2}d)\@. 

\section{Analsyis of the Sn-Pc-$u$ monomer electrofluorescence}

\begin{figure}
\centering
\includegraphics[width=\textwidth]{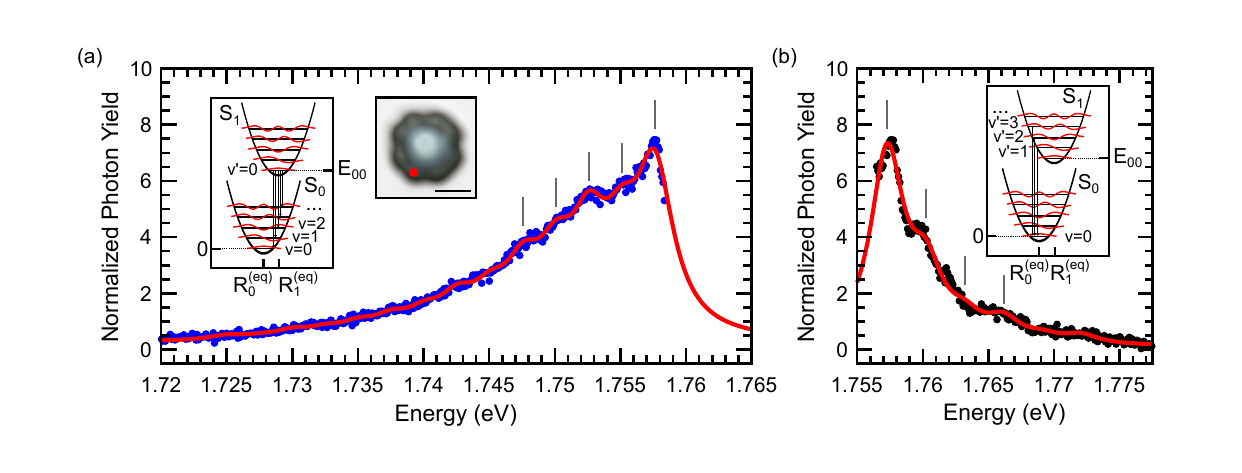}
\caption{Line shape analysis of the Sn-Pc-$u$ photon yield spectrum.
(a) Low-energy ($E\leq E_{00}$) side of STML spectrum (dots) measured with the tip placed above an isoindole moiety of Sn-Pc-$u$ ($1.95\,\text{V}$, $50\,\text{pA}$, $180\,\text{s}$)\@. 
The spectrum is normalized by the NCP electroluminescence spectrum ($1.9\,\text{V}$, $50\,\text{pA}$, $120\,\text{s}$) above bare NaCl. 
The solid line represents a fit to the data with equidistantly spaced Lorentzians and identical width. 
Insets: illustration of vibrational progression (left) and constant-height current map of Sn-Pc-$u$ (right, $1.9\,\text{V}$, the gray scale encodes currents ranging from $10$ (dark) to $100\,\text{pA}$ (bright), scale bar: $1\,\text{nm}$)\@.
(b) As (a) for high-energy ($E\geq E_{00}$) side of the STML spectrum.
Inset: illustration of $S_1\rightarrow S_0$ transitions for hot luminescence.}
\label{figS3}
\end{figure}

Figure \ref{figS3} investigates the Sn-Pc-$u$ emission in detail. 
The visible spectroscopic fine structure is assigned to vibronic transitions. 
Each of them represents the relaxation from the vibrational state $\nu'$ of the excited electronic state $S_1$ to the vibrational state $\nu$ of $S_0$ (insets to Figures \ref{figS3}a,b)\@. 
Within the Franck-Condon picture, the individual peak heights are defined by $|\braket{\chi_\nu|{\chi_{\nu'}}}|^2$ ($\chi_{\nu},\chi_{\nu'}$: wave functions of vibrational states $\nu$, $\nu'$) weighted by the occupation $p_{\nu'}$ of level $\nu'$. 
Close to the vibrational equilibrium coordinates $R_{0}^{(\text{eq})}$ and $R_{1}^{(\text{eq})}$, the harmonic potentials $V_j\propto k_j\cdot\left[R-R_j^{(\text{eq})}\right]^2$ ($j=0,1$) with the generalized coordinate $R$ describe well the potential landscape for vibrations in the electronic states $S_0$ and $S_1$\@.
Generally speaking, the vibrational potentials for $S_0$ and $S_1$ do not necessarily have the same curvature. For example, the restoring forces, which in the harmonic approximation are proportional to $k_j$, for frustrated rotations of metal-phthalocyanines on a NaCl substrate were demonstrated to differ for the electronic ground and excited state \cite{natcommun_13_6008}. 

In the case $R_0^{(\text{eq})}\approx R_1^{(\text{eq})}$ and $k_0\approx k_1$, the Franck-Condon factor prefers transitions with $\nu=\nu'$.
According to Kasha's rule \cite{faradaysoc_9_14}, emission occurs predominantly from the $\nu'=0$ level of $S_1$\@.
Therefore, the strongest photon yield observed at $1.758\,\text{eV}$ (Figures \ref{figS3}a,b) is attributed to the principal transition between the vibrational ground states of $S_0$ and $S_1$ with energy $E_{\nu\nu'} = E_{00}$\@. 
The low-energy tail ($E\leq E_{00}$) of the luminescence spectrum (Figure \ref{figS3}a) contains the transitions from $\nu'=0$ to $\nu>\nu'$ with an energy separation given by the vibrational energy quantum $\varepsilon_0$ of $S_0$\@.

Matching the low-energy tail of the Sn-Pc-$u$ principal emission peak with a superposition of Lorentzians with equidistant energy spacing (solid line in Figure \ref{figS3}a) allows to extract $\varepsilon_0$. 
The broadening of the emission peaks due to transitions from $\nu'>0$ as well as due to the finite lifetime of $S_1$ and experimental resolution is accounted for by a finite Lorentzian width, which is chose identical for all Lorentzians in the fit.
As a result of the fit, $\varepsilon_0=2.51\pm0.03\,\text{meV}$\@.  
The uncertainty margin reflects the $95\,\%$ confidence interval of the fit parameter $\varepsilon_0$ underlying the least-squares fit to the electrofluorescence spectrum.

For the emission fine structure in the high-energy tail of the Sn-Pc-$u$ principal transition, vibrational excited state of $S_1$ ($\nu'>0$) have to be considered (inset to Figure \ref{figS3}b)\@. 
The underlying hot luminescence requires $p_{\nu'} \neq 0$ for $\nu'>0$\@. 
Apparently, direct tunneling into excited vibrational states and the enhanced radiative decay rate due to the NCP violates Kasha's rule \cite{faradaysoc_9_14}, as reported previously \cite{natcommun_13_6008,nl_16_6480,nl_18_3407,nl_21_7086}.   
The empirical Kasha rule \cite{faradaysoc_9_14} forbids molecular luminescence due to nonthermalized excitons.
However, the fluorescence lifetime in a plasmonic environment can significantly be reduced and become comparable with the vibrational relaxation time in an electronic state.
Therefore, luminescence peaks blueshifted from the principal $S_1(\nu'=0)\rightarrow S_0(\nu=0)$ transition can indeed occur.
The energy spacing for the hot luminescence features is determined by $\varepsilon_1$ of $S_1$, which again results from a fit to the data using a superposition of Lorentzians (solid line in Figure \ref{figS2}b) as $\varepsilon_1=2.96\pm0.07\,\text{meV}$\@.

Because of the presence of hot luminescence, $p_{\nu'}\neq 0$ for $\nu'\geq 1$\@.
Therefore, vibrational progression can occur from these vibrationally excited states of $S_1$, too, which leads to a broadening of the vibrational progression peaks in the series with $\nu'=0$\@.
For example, the vibrational progression transition $S_1(\nu'=1)\rightarrow S_0(\nu=2)$ exhibits the energy $E_{21}=E_{10}+\varepsilon_1-\varepsilon_0$, which is very close to the $S_1(\nu'=0)\rightarrow S_0(\nu=1)$ transition with energy $E_{10}$\@.

\section{Spectroscopy of Sn-Pc monomer and dimer orbitals}

\begin{figure}
\centering
\includegraphics[width=\textwidth]{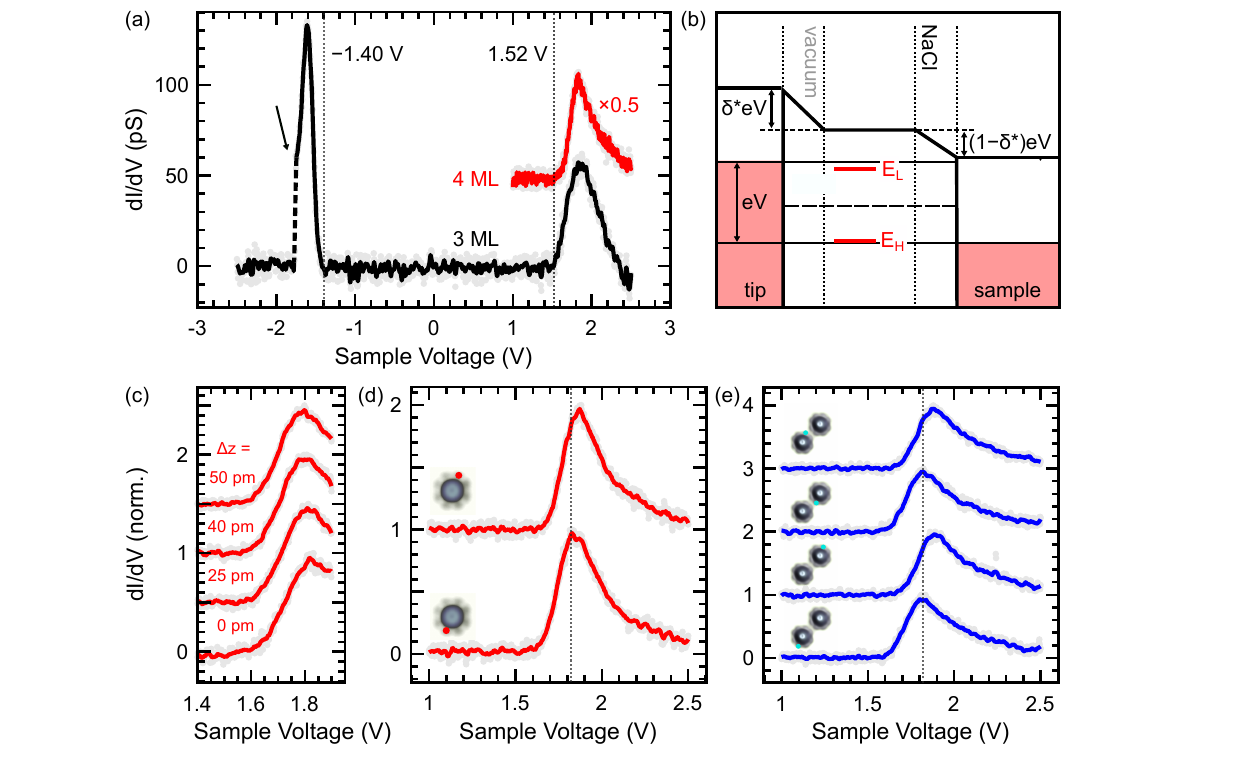}
\caption{Spectroscopy of Sn-Pc-$u$ monomer and dimer orbitals.
(a) Wide-range spectrum (dots) of $\text{d}I/\text{d}V$ of Sn-Pc-$u$ on $3$ atomic layers of NaCl on Au(111)\@. 
The signature related to the HOMO (LUMO) appears as a peak with an onset at $V_{\text{H}}=-1.40\,\text{V}$ ($V_{\text{L}}=1.52\,\text{V}$)\@.
The black arrow marks the abrupt drop of the spectroscopic signal to $0$ caused by the unintentional lateral displacement of the molecule. 
The LUMO feature of Sn-Pc-$u$ adsorbed on a $4$-layer NaCl island is vertically offset. 
Feedback loop parameters of both spectra: $1.9\,\text{V}$, $20\,\text{pA}$\@. 
(b) Illustration of the double tunneling barrier across the vacuum gap and the NaCl film (see text)\@.
(c) Dependence of the $\text{d}I/\text{d}V$ LUMO-related signature on tip approach (from bottom to top)\@.
The feedback loop parameters $1.9\,\text{V}$, $20\,\text{pA}$ define the tip displacement $\Delta z=0$\@.
(d) Dependence of the $\text{d}I/\text{d}V$ LUMO-related signature ($1.9\,\text{V}$, $20\,\text{pA}$) on the intramolecular site. 
The dashed line marks the peak position of the lower spectrum. 
Insets: constant-height current maps of the Sn-Pc-$u$ monomer with indicated spectroscopy sites. 
(e) Same as (d) for an $uu$ dimer.
In (c)--(d), the spectra are normalized to their respective maximum and are vertically offset.}
\label{figS4}
\end{figure}

Figure \ref{figS4} shows spectra of $\text{d}I/\text{d}V$ obtained from Sn-Pc-$u$ monomers and dimers.  
The wide-range spectrum in Figure \ref{figS4}a was acquired above the center of Sn-Pc-$u$ on a $3$-layer NaCl island.
The onsets of peaks at negative ($V_{\text{H}}$) and positive ($V_{\text{L}}$) voltage (dotted lines in Figure \ref{figS4}a) are related to the HOMO and LUMO of Sn-Pc-$u$, respectively.
To estimate the orbital energies $E_{\text{H}}$ (HOMO) and $E_{\text{L}}$ (LUMO), the voltage drop across the vacuum barrier ($\delta V$) and across the NaCl film ($(1-\delta)V$) must be considered (Figure \ref{figS4}b) as well as the difference between the HOMO (LUMO) and the measured positive (negative) ion resonance \cite{jpcm_20_184001}.
In the following, both effects are combined in a single factor $\delta^\ast$, i.\,e., $E_{\text{H}}=\delta^\ast\text{e}V_{\text{H}}$ and $E_{\text{L}}=\delta^\ast\text{e}V_{\text{L}}$ (e: elementary charge), which is in accordance with previous reports \cite{prl_93_236802,prl_112_047403,apl_107_043103,nature_531_623}.
For Sn-Pc-$u$ on a $3$-layer NaCl island, $V_{\text{H}}=-1.40\,\text{V}$ and $V_{\text{L}}=1.52\,\text{V}$ (Figure \ref{figS4}a)\@.
The principal $S_1\rightarrow S_0$ transition of Sn-Pc-$u$ on a $3$-layer NaCl island is observed at the energy $E_{00}=1.76\,\text{eV}$ (not shown)\@.
Assuming now that the HOMO-LUMO gap approximately equals the Q-exciton energy \cite{nature_531_623, natcommun_14_4988}, i.\,e., $E_{00}\approx E_{\text{L}}-E_{\text{H}}=\delta^\ast\text{e}(V_{\text{L}}-V_{\text{H}})$, yields $\delta^\ast=0.60$, $E_{\text{L}}=0.92\,\text{eV}$, and $E_{\text{H}}=-0.84\,\text{eV}$ for Sn-Pc-$u$ on a $3$-layer NaCl island.
For Sn-Pc-$u$ on a $4$-layer NaCl island the onset of the LUMO signature (vertically offset data in Figure \ref{figS4}a, recorded above an isoindole group) appears at $V_{\text{L}}=1.63\,\text{V}$ for the same feedback loop parameters as in the case of the $3$-layer NaCl island, which according to $E_{\text{L}}=\delta^\ast\text{e}V_{\text{L}}$ results in $\delta^\ast=0.56$\@. 
The HOMO signature was not accessible on $4$-layer NaCl due to junction instabilities at elevated negative voltages.

Figures 2b,e of the article show that the maximum total photon yield of Sn-Pc-$u$ on a $4$-layer NaCl island is reached at about $\hat{V}=2\,\text{V}$\@. 
With the estimated $\delta^\ast=0.56$, the energy level alignment for the optimal photon yield can be derived. 
The chemical potential of the tip exhibits the energy distance $\delta^\ast\text{e}\hat{V}-E_{\text{L}}=0.2\,\text{eV}$ above the LUMO level, while the chemical potential of the sample is $E_{\text{H}}-(1-\delta^\ast)\text{e}\hat{V}=0.06\,\text{eV}$ above the HOMO level. 
This level alignment at $\hat{V}$ is schematically depicted in Figure \ref{figS4}b. 

Figure \ref{figS4}c depicts the dependence of the LUMO-associated $\text{d}I/\text{d}V$ spectrum of Sn-Pc-$u$ on the tip approach by $\Delta z$ to the center of an isoindole moiety. 
Due to the voltage drop across the vacuum barrier, a shift of the LUMO-related signature to higher voltages would be expected. 
However, the signature shifts to lower voltages by about $30\,\text{mV}$. 
Possibly, the voltage drop is overcompensated by the Stark effect \cite{prl_91_196801,prb_70_033401} the orbital is subject to in the electric field between the tip and the sample.
The LUMO-related $\text{d}I/\text{d}V$ signal likewise shows little variation ($<100\,\text{meV}$) across different lobes both for the monomer and the dimer (Figures \ref{figS4}d,e), which is tentatively assigned to local electrostatic variations of the underlying ionic NaCl lattice. 
Supported by the unaltered emission spectra across all lobes of the monomer this small variation in the LUMO energy is irrelevant for the light emission of Sn-Pc-$u$ monomers and dimers. 

\section{Fabrication of dimers and control of configurational mo\-lecular states} 

\begin{figure}
\centering
\includegraphics[width=\textwidth]{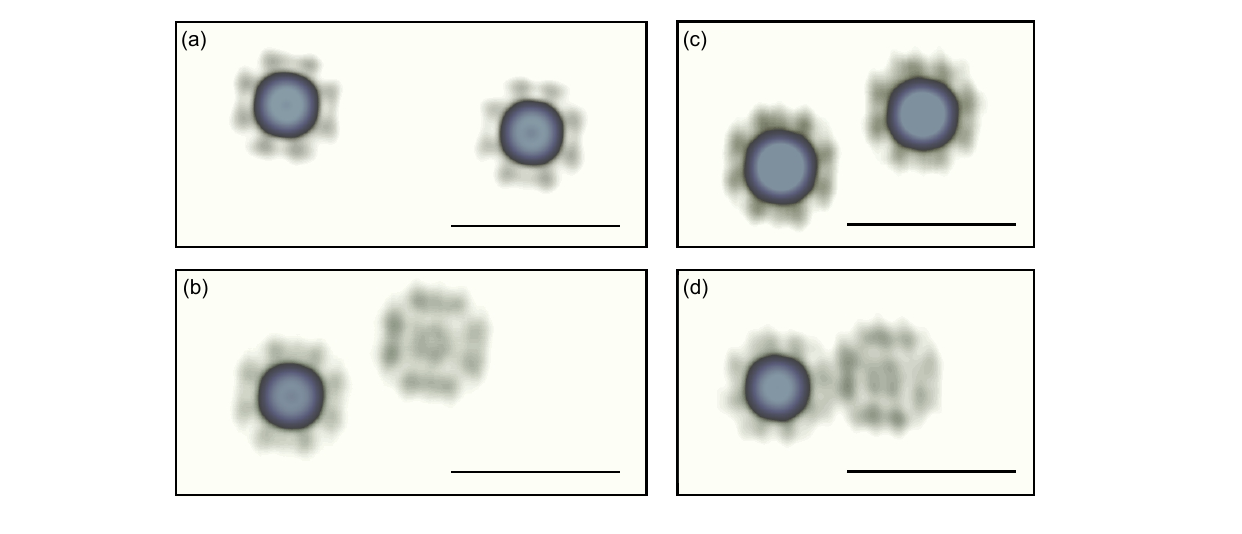}
\caption{Lateral and configurational manipulation of Sn-Pc.
(a) Constant-height current map of two separated Sn-Pc-$u$ molecules on $4$ atomic layers of NaCl ($1.9\,\text{V}$, the gray scale encodes currents ranging from $2$ (white) to $30\,\text{pA}$ (gray))\@. 
(b) Same as (a) after the first lateral manipulation step applied to the right Sn-Pc-$u$ in (a)\@.
A $u\rightarrow d$ conversion occurred to the manipulated Sn-Pc.
(c) Same as (a) after conversion of the right Sn-Pc-$d$ in (b) back to Sn-Pc-$u$.
(d) As (a) after the second lateral manipulation step applied to the right Sn-Pc-$u$ in (c)\@.
The scale bar in (a)--(d) depicts $3\,\text{nm}$.}
\label{figS5}
\end{figure}

Two separated Sn-Pc-$u$ molecules (Figure \ref{figS5}a) are pushed toward each other by placing the tip atop an isoindole group with feedback loop parameters $-3.1\,\text{V}$ and $20\,\text{pA}$\@.
A successful pushing event of the molecule is reflected by a sudden current drop. 
A subsequent STM image of the same area (Figure \ref{figS5}b) shows the manipulated molecule which during the manipulation process switched to Sn-Pc-$d$. 
The $d\rightarrow u$ conversion is achieved by placing the tip above the Sn atom at positive sample voltage ($1.9\,\text{V}$) with enabled feedback loop. 
Gradually increasing the tunneling current reliably leads to a sudden increase of the tip--sample distance, which signals the changed position of the Sn atom (Figure \ref{figS5}c)\@. 
The dimer fabrication is completed by executing another pushing cycle to decrease the mutual distance (Figure \ref{figS5}d)\@. 

The $u\rightarrow d$ conversion can be induced without moving the molecule when the tip is parked above the center of the molecule. 
After deactivating the feedback loop at $1.9\,\text{V}$, $5\,\text{pA}$ the tip is retracted $60\,\text{pm}$.
By gradually decreasing the sample voltage to $\approx-2.7\,\text{V}$ gives rise to a sudden current drop, which signals successful conversion of Sn-Pc-$u$ to Sn-Pc-$d$. 

\section{Dependence of the dimer emission on the mutual mo\-no\-mer separation}

\begin{figure}
\centering
\includegraphics[width=\textwidth]{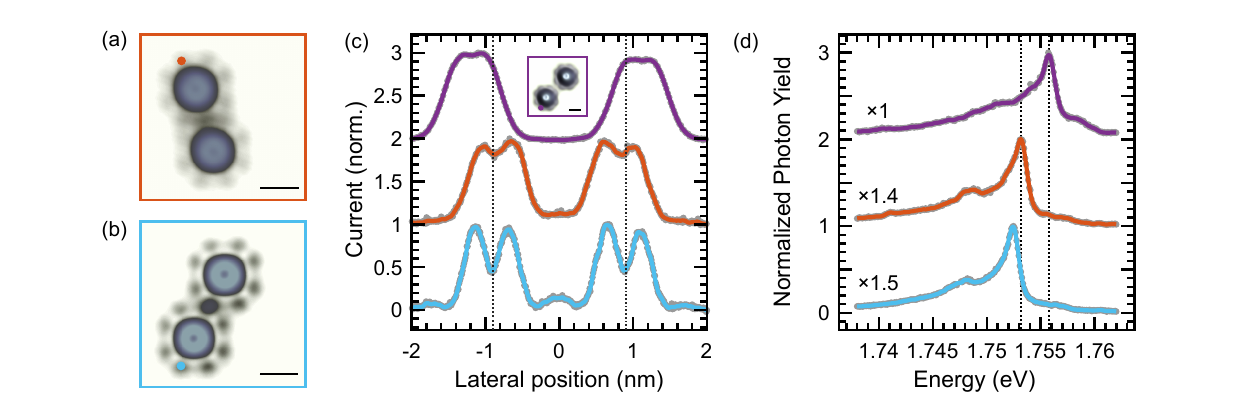}
\caption{Dependence of the Sn-Pc-$u$ dimer emission on the mutual monomer separation.
(a) Constant-height STM image of an $uu$ dimer with center-to-center distance of monomers of $1.7\,\text{nm}$ ($1.9\,\text{V}$, the gray scale encodes currents between $1.5$ (dark) and $30\,\text{pA}$ (bright))\@.
(b) Constant-height STM image of an $uu$ dimer with center-to-center distance of monomers of $1.8\,\text{nm}$ ($1.9\, \text{V}$, the gray scale encodes currents between $1.5$ (dark) and $30\,\text{pA}$ (bright))\@.
(c) Cross-sectional profiles of constant-height current maps acquired atop $uu$ dimers along a line connecting the molecular centers with mutual distances of $2.3\,\text{nm}$ (top), $1.7\,\text{nm}$ (middle), $1.8\,\text{nm}$ (bottom)\@.
(d) Comparison of normalized photon yield ($1.9\,\text{V}$, $50\,\text{pA}$, $240\,\text{s}$ (top);  $1.9\,\text{V}$, $50\,\text{pA}$, $180\,\text{s}$ (middle); $1.9\,\text{V}$, $50\,\text{pA}$, $180\,\text{s}$ (bottom)) of the $uu$ dimers in (c)\@.
The spectra are normalized by the NCP luminescence spectrum acquired atop bare NaCl as well as to their respective maximum and are vertically offset.
In (a),(b) and the inset to (c) the scale bar depicts $1\,\text{nm}$.}
\label{figS6}
\end{figure}

The $uu$ dimer luminescence spectrum exhibits an overall redshift as a function of the separation of its constituent monomers (Figure \ref{figS6})\@.
According to the point-dipole approximation this observation agrees well with expectations because $J_0\propto 1/r^{3}$ (see article text)\@. 
It is noteworthy in this context that for dimers with similar center-to-center distances of the monomers, the actual geometry of the monomer-monomer hybridization is important.
The split lobes of the monomers can be interlocked, that is, a lobe of one monomer is positioned between two adjacent lobes of the other monomer (Figure \ref{figS6}a)\@.
In another configuration the lobes overlap (Figure \ref{figS6}b)\@.
The two bottom luminescence spectra of Figure \ref{figS6}d show that the overlap configuration gives rise to a redshift of photon energies compared to the interlock configuration of the $uu$ dimer. 

\section{Adsorption of Sn-Pc on ultrathin NaCl films}

\begin{figure}
\centering
\includegraphics[width=\textwidth]{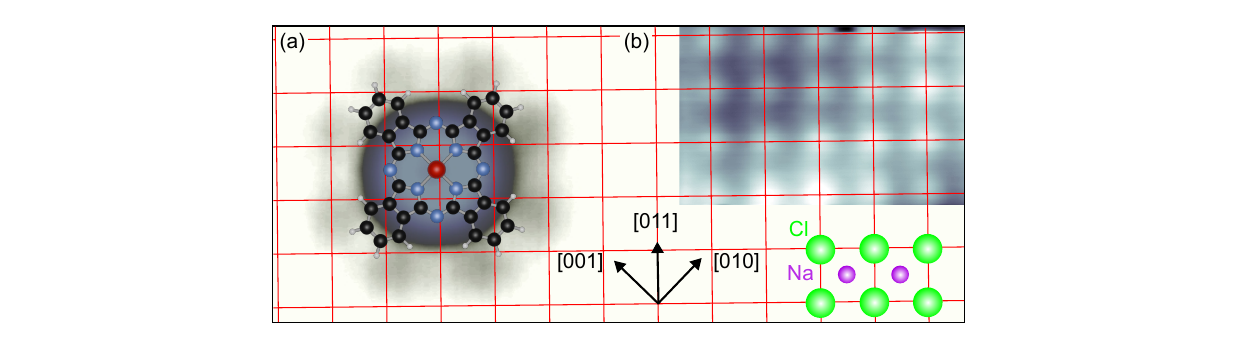}
\caption{Adsorption of Sn-Pc-$U$ on NaCl.
(a) Constant-height STM image of an Sn-Pc-$u$ adsorbed on a $4$-layer NaCl island ($1.9\,\text{V}$, the gray scale encodes currents between $2\,\text{pA}$ (dark) and $40\,\text{pA}$ (bright))\@.
The superimposed molecule model reflects the relaxed Sn-Pc vacuum structure.
(b) Constant-current STM image of the nearby NaCl lattice ($1.9\,\text{V}$, $35\,\text{pA}$, the NaCl lattice has a corrugation of $4\,\text{pm}$)\@.
The grid lines intersect at the centers of Cl ions (circular protrusions)\@.
Compact directions ($[010]$, $[001]$, $[011]$) and a sketch of the NaCl lattice are shown.}
\label{figS7}
\end{figure}

For Sn-Pc-$u$ on a $4$-layer NaCl island (Figure \ref{figS7}a) the closely neighboring NaCl lattice was imaged with atomic resolution (Figure \ref{figS7}b).
Extending the position of Cl ions, which are clearly discerned as circular protrusions \cite{surfsci_424_321,jpcm_24_475507,nature_531_623, acsnano_13_6947}, to the Sn-Pc-$u$ adsorption site allows to analyze the adsorption site in more detail.
The Sn atom adsorbs on a bridge site between two Cl ions, while the lobes point along the NaCl compact directions $[001]$ and $[010]$\@. 
The lobes at the periphery of the isoindole groups cover inequivalent areas of the NaCl lattice.

%\bibliography{ref.bib}
\providecommand{\latin}[1]{#1}
\makeatletter
\providecommand{\doi}
  {\begingroup\let\do\@makeother\dospecials
  \catcode`\{=1 \catcode`\}=2 \doi@aux}
\providecommand{\doi@aux}[1]{\endgroup\texttt{#1}}
\makeatother
\providecommand*\mcitethebibliography{\thebibliography}
\csname @ifundefined\endcsname{endmcitethebibliography}
  {\let\endmcitethebibliography\endthebibliography}{}